# Simulations of GRB detections with the ECLAIRs telescope onboard the future SVOM mission

___________________________________________________________________________

**Sarah Antier[1]\*, Stéphane Schanne[1], Bertrand Cordier[1], Aleksandra Gros[1], Diego Götz[1], Cyril Lachaud[2]**

[1]*CEA Saclay, IRFU/Service d'Astrophysique, 91191 Gif sur Yvette, France*
[2]*APC, UMR 7164, 10 rue Alice Domon et Léonie Duquet 75205 Paris, France*

*E-mail:* `sarah.antier@cea.fr`

The soft gamma-ray telescope ECLAIRs with its Scientific Trigger Unit is in charge of detecting Gamma-Ray Bursts (GRBs) on-board the future SVOM satellite. Using the "scientific software model" (SSM), we study the efficiency of both implemented trigger algorithms, the *Count-Rate Trigger* for time-scales below 20 s and the *Image Trigger* for larger ones. The SMM provides a simulation of ECLAIRs with photon projection through the coded-mask onto the detection plane. We developed an input GRB database for the SSM based on GRBs light curves detected by the Fermi GBM instrument. We extrapolated the GRB spectra into the ECLAIRs band (4-120 keV) and projected them onto the detection plane, superimposed with cosmic extragalactic background photons (CXB). Several simulations were performed by varying the GRB properties (fluxes and positions in the field of view). We present first results of this study in this paper.

___________________________________________________________________________

[*]Speaker

1.     **The Gamma-Ray Burst trigger telescope ECLAIRs onboard SVOM**

Gamma-Ray Bursts (GRBs), the most energetic flashes of γ-rays, continue to capture the attention of the scientific community since their discovery in the late 1960's. GRBs are characterized by a transient gamma-ray prompt emission, which appears at a random location on the sky, followed by a long-lived, multi-wavelength afterglow emission (X-ray, ultraviolet, optical, infrared, microwave and radio). Thanks to several past and present space missions such as CGRO [1], INTEGRAL [2], Swift and Fermi [3], GRBs science is an active field of research.

The GRB prompt emission in gamma-rays is believed to be created by internal shocks in a jet of matter produced at cosmological distances. It comes from the formation of a black hole during stellar collapse or neutron-star merger. GRBs continue to raise questions in physics, such as the involved prompt-emission processes or the nature of ultra-longs GRBs [4].

The SVOM mission (Space-based Variable Objects Monitor) [1] is a Chinese-French multi-wavelength and wide field-of-view (FOV) observatory expected to be launched in 2021. The main objective of the mission is the detection and the multi-wavelength follow-up of transient sources such as GRBs in space and on ground. SVOM has four instruments on board and three instruments on ground. The space instrumentation includes the GRB trigger telescope ECLAIRs [6] and the Micro-channel X-ray Telescope MXT (both provided by France), as well as the Gamma-Ray Monitor GRM and the Visible Telescope VT.

ECLAIRs is composed by a wide FOV (2 sr) soft-gamma ray telescope (TXG) and its associated Scientific Trigger Unit (UGTS). The TXG energy range is 4-150 keV, its detection area is 1024 cm$^2$. A coded mask of 40% aperture is placed 46 cm above the detector plane. Its dimensions are 54×54 cm$^2$, and the ratio *m/d* between the element size of the mask (*m*) and the detector (*d*) is 2.6. The UGTS (Control & Scientific Trigger Unit of ECLAIRs) is in charge of the command/control of the camera, the data acquisition and the near real-time data processing by two concurrent GRB triggers algorithms, the *Count-Rate Trigger* and the *Image Trigger*, providing fast detections and localizations of GRBs. The generated GRB alerts are transmitted to the spacecraft for autonomous repointing and to ground via a VHF network.

2.     **Scientific Software Model : an implementation of the trigger algorithms**

Both trigger algorithms, the *Count-Rate Trigger* and the *Image Trigger*, foreseen for the flight hardware, can be compiled in the so-called *Scientific Software Model* [7] on a much faster standard linux computer, which allows to test the scientific performances of the algorithms.

2.1    **Image Trigger**

The *Image Trigger* algorithm has been implemented to detect GRBs of durations above tens of seconds. It runs on different time-scales from 20.48 s to ~20 min in 4 different energy bands. Every 20.48 s, the background in the shadowgram (accumulated counts per pixel) is modelled by fitting a 2D 2$^{nd}$ order polynomial function. The background subtracted shadowgram is deconvolved using the mask pattern to obtain sky images (in counts and variance). Sky images of longer time-scales are built by summation of shorter ones. In each sky image built, pixels not masked by Earth have their Signal to Noise Ratio (*SNRimage=counts/√variance*) evaluated. The localization of the best excess exceeding a given *SNRimage*-threshold, which does not correspond to a known source position, identifies a new

GRB source. In order to obtain a finer localization than the sky binning, the excess peak in the counts image is fit by a 2D Gaussian function with width fixed by the ratio *m/d*.

## 2.2   Count-Rate Trigger

The *Count-Rate Trigger* algorithm has been implemented to detect GRBs of durations from 10 ms to ~20 s. It is divided into two main parts: the calculation of count-rate excesses and the determination of the best excess to be imaged for new source detection and localization.

Count-rate excesses are detected on different logarithmic time-scales from 10 ms to 20.48 s, in four different energy bands and on nine overlapping detector zones (full detector, four halves and four quadrants). A background-count estimate for each time-scale is computed by extrapolating the counts from previous time-scales. The SNR of each time-scale is calculated from the number of counts in the time-scale ($N$) and the estimated background counts ($B$) as: $SNR = (N-B)/\sqrt{B}$. If the SNR of a time-scale exceeds a threshold, an excess is detected and stored in a buffer. Such an excess is characterized by its time-scale, energy band and zone. Every 2.56 s the best excess stored in the buffer and which is not too old (40 s into the past) is searched and a sky-image using the photons detected in its time-scale and energy band is constructed by building their shadowgram and reconstructing the corresponding sky image. In this image, excesses above a given *SNRimage*-threshold are searched and if they correspond to a new source, they are fine-localized in a procedure identical to the one of the *Image Trigger*.

## 3.   Simulation of ECLAIRs Trigger Performances

The *Scientific Software Model* input is a list of photons (from GRBs, superimposed on background) characterized by their time and energy. The output is a sequence detected GRBs.

### 3.1   GRB input data base

A GRB input-database has been created from real GRBs, detected by the Gamma-Ray Burst Monitor (GBM) onboard the Fermi Gamma-Ray Space Telescope, from June 2008 to July 2012 (819 GRBs). The GBM spectral catalogue [8] gives access to spectral parameters for each GRB through their best-fit model (power-law, Comptonized model, Band model, smoothly-broken power-law), obtained in the GBM sensitive energy range (8-1000 keV) after background noise removal. The corresponding light-curve obtained after background subtraction is also accessible [9], with a minimal time-resolution of 64 ms.

For each GBM burst, using the spectral model and corresponding parameters, the spectrum is converted into the 4-120 keV energy range of ECLAIRs, a small extrapolation of the spectrum being performed from the 8 keV low-energy threshold of GBM down to the 4 keV one of ECLAIRs. From the time-integrated spectrum the total flux (in ph/cm$^2$/s) is calculated.

Using the duration of the burst and the ECLAIRs detector area, the number $N$ of GRB photons impinging on axis on the ECLAIRs detector in its energy range is computed. A list of GRB photons is then created, containing $N$ photons characterized by a time determined within the GRB duration and an energy determined in the 4-120 keV range. To compute the photon times, the cumulative time-distribution is constructed: each time-bin (64 ms minimum) contains the integral of the light-curve from the start of the burst to the current time-bin. The cumulative time-distribution (Figure 1a) permits a bijection. After random uniform draws on the y-axis, we determine the corresponding time on the x-axis for each of the $N$ photons. The same procedure is applied with the cumulative spectrum to associate an energy to each photon. An example of a simulated GRB in the ECLAIRs energy range is shown on Figure 1b.

### 3.2    Simulations setup

For the simulations presented here we generate one simple background-photon file, covering 1000 s, without Earth transits in the FOV of ECLAIRs. Those photons follow the Cosmic X-ray Background (CXB) energy spectrum [10] and arrive with isotropic distribution inside the ECLAIRs FOV. We generate also different sets of GRB photon-files from GBM data as described before. In one set the GRBs cover the 4-120 keV ECLAIRs energy range, in another set they cover the reduced energy range 15-120 keV. From each of those sets, weaker GRBs are generated by taking 1 out of $F$ photons ($F$ ranging from 1 to 20) from the GRB photon-list, which artificially reduces their flux by a factor $F$ while preserving their timing and spectral properties. In a first simulation run the GRBs are all placed on axis, in a second run they are randomly placed (with isotropic distribution) inside the ECLAIRs FOV. During the simulation, each photon from the input list, resulting from the merger of the GRB photon and background photon lists, is projected by ray-tracing through the ECLAIRs model. It takes into account the geometry of the mask and the detector, its properties [11] (detector efficiency, mask transparency) and its simulated instrumental background.

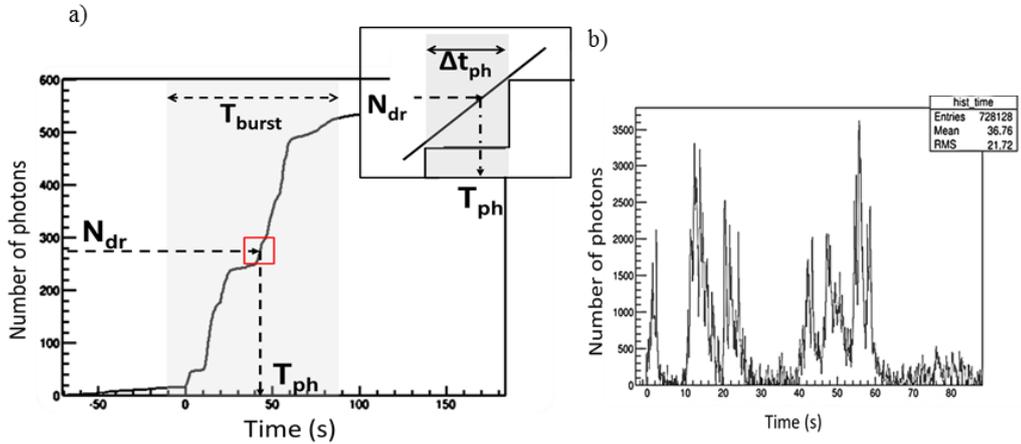

Figure 1. a) Cumulative time-distribution of GRB 080723B to display the way to derive a time for each simulated photon. b) Light curve of simulated GRB 080723B in the ECLAIRs energy range.

### 3.3    Result of ECLAIRs Trigger performances

#### 3.3.1    Global performance results

In the simulations we obtain a 99.9% detection efficiency for simulated on-axis GBM bursts (1st run) and 98.0% ±0.6% for bursts randomly placed in the ECLAIRs FOV (mean and standard deviation determined by repeating the 2nd run 4 times). These results are obtained for the GBM bursts in the ECLAIRs energy range without reduction factor ($F$=1), which are quite intense bursts, with the mean {and standard deviation} of their logarithmic flux distribution (log-mean flux hereafter) being $Log_{10}(\phi$ [ph/cm$^2$/s]$) = 3.3\{2.1\}$. For weaker bursts ($F = 8$), their log-mean flux is 0.41{2.1} in the same energy range, 80.0 % detection efficiency is obtained on-axis, 57.4 % ± 1.3% when random in the FOV. No false detections have been recorded with our simple background model used, the *SNRimage*-threshold being set to 6.5σ in all our simulations.

Figure 2 shows the fraction of simulated GBM bursts detected by at least one trigger algorithm (*Count-Rate Trigger* or *Image Trigger*) as a function of their flux reduction factor *F*. As the reduction factor *F* increases, the detection efficiencies decrease and the relative difference increases between on-axis and random positions. For example with *F*=2 the difference is 2% ± 0.5%, whereas for *F*=8 it is 28% ± 2%. A previous study [7] of ECLAIRs trigger-performances was performed with the CXB background modulated by Earth transits in the FOV, with simulated BATSE bursts extrapolated into the ECLAIRs energy range, randomly placed in the part of the FOV non-obscured by the Earth. In the 4-120 keV energy range, the log-mean flux of BATSE bursts [12] is 1.37{2.75}. A 88.4% detection efficiency was reported. The mask ratio *m/d* was set to 1.2 in this study, and meanwhile updated to 2.6 to increase burst detections [6].

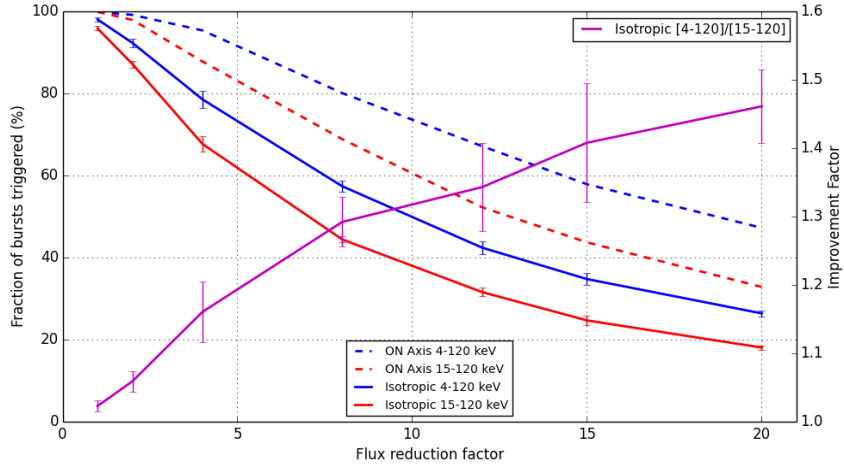

Figure 2. Detection performances, by at least one of both trigger algorithms using simulated GBM bursts in the standard or reduced energy band for on-axis and randomly distributed bursts, vs flux reduction F. The improvement factor using a 4 keV instead of 15 keV low-energy threshold is shown in the case of simulated GBM bursts randomly distributed in the ECLAIRs FOV.

### 3.3.2   Comparison between Count-Rate Trigger and Image Trigger

The detection efficiency is 90.8% ± 1.3% for the *Image Trigger* and 97.2% ± 0.90% for the *Count-Rate Trigger* on simulated GBM bursts (*F*=1) randomly placed in the ECLAIRs FOV and with standard CXB background without Earth transits, nor additional persistent sources. Most bursts are detected by both algorithms. Figure 3 shows the detection fraction of simulated GBM bursts as a function of the flux reduction factor *F* for the *Count-Rate Trigger* and the *Image Trigger*. For intense simulated bursts (*F*=1), the *Count-Rate Trigger* is more efficient than the *Image Trigger*. Indeed, the *Image Trigger* mostly does not detect short bursts with low fluence. For *F*=1, 66%±4.0% of short bursts are detected by the *Image Trigger*, compared to 96%±1.2% for the *Count-Rate Trigger*. However, the *Image Trigger* recovers bursts with low peak-flux, not detected by the *Count-Rate Trigger*. For example for *F*=8, the log-mean peak-flux of GRBs detected by the *Image Trigger* is 0.92 {2.8}, while it is 1.1 {2.6} for the *Count-Rate Trigger*.

### 3.3.3   Influence of the 4 keV low-energy threshold

We compare the detection performances of the trigger in the standard ECLAIRs energy range to the reduced 15-150 keV energy range (see Figure 2). We observe that standard GBM bursts (GRBs with high fluxes) are detected in both energy ranges. However, the threshold of 4 keV permits a net increase of 30% ± 5% of weak GRBs detections (*F*=8), despite the higher

CXB flux in the 4-15 energy range. The inscrease is observed both for the *Count-Rate* and the *Image Trigger*. For example for *F*=8, the log-mean flux of GBM bursts detected by both triggers in 4-120 keV is 0.52 {2.0}, whereas it is 0.55 {2.0} in 15-120 keV.

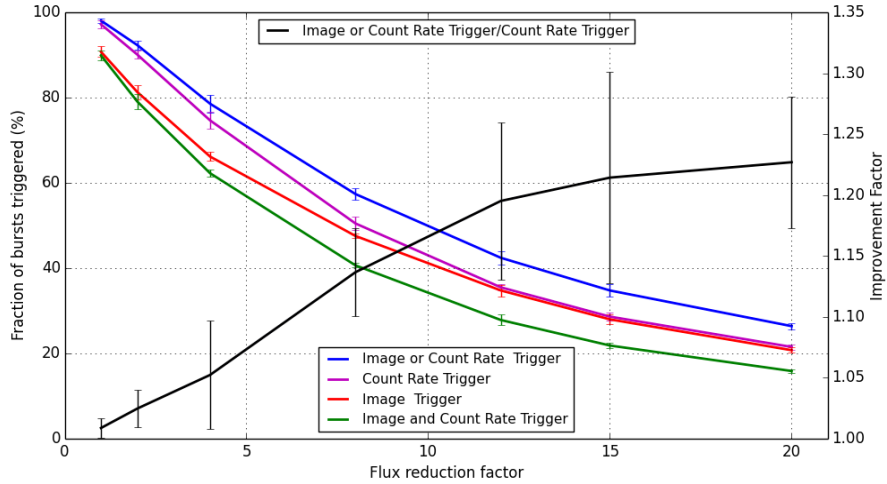

Figure 3. Comparison of the *Count-Rate* and *Image Trigger* detection performance in the 4-120 keV energy band for randomly distributed bursts, vs flux reduction *F*. The curves for bursts detected by the two algorithms are shown, as well as the improvement using both algorithms concurrently compared to the *Count-Rate Trigger* alone.

## 4.     Conclusion

In this paper, we presented first results on the scientific-performance studies of the GRB detections with the ECLAIRs telescope onboard SVOM. Detection efficiencies have been evaluated using an ECLAIRs Monte-Carlo model and the two trigger algorithms (*Count-Rate Trigger* and *Image Trigger*). We built an imput data-base of 819 simulated GBM bursts with light-curves and spectra extrapolated into the ECLAIRs energy band (4-120 keV) overlaid with simple CXB background without Earth transits in the FOV. The results show that the trigger algorithms detects almost all simulated GBM bursts (98.0% ±0.6%) randomly placed in the FOV. Short GRBs are preferentially found by the *Count-Rate Trigger*. Weak GRBs are preferentially found by the *Image Trigger*. A low-energy threshold of 4 keV instead of 15 keV permits to detect weaker simulated GBM bursts.

## References

[1]     G.J. Fishman & C.A. Meegan, *Gamma-Ray Bursts*, *Annual Review of Astronomy and Astrophysics*, **33** (1995) 415-458

[2]     D.Götz et al., *INTEGRAL Results on Gamma-Ray Bursts,POS Integral*, **117** (2013)

[3]     N.Gehrels & S.Razzaque, *Gamma-ray bursts in the Swift-Fermi era, Frontiers of Physics*, **8** (2013) 671-678

[4]     P. Kumar, B. Zhang, *The physics of gamma-ray bursts & relativic jet*, *Physics Report* (2015)

[5]     B. Cordier et al., *The Chinese-French SVOM Mission for gamma-ray burst study, Proc. Swift 10 years of discovery* (2015)

[6]     S. Schanne et al., ECLAIRs: *The gamma-ray imager of the SVOM satellite*, *Proc. Swift 10 years of discovery* (2015)

[7]     S. Schanne et al., *A Scientific Trigger Unit for Space-Base Real-Time Gamma Burst Detection, Proc. IEEE NSS* (2013)

[8]   D. Gruber et al., *The Fermi GBM Gamma-Ray Burst Spectral Catalog : Four Years of Data* , *ApJs*, **211** (2009) 1

[9]   A. von Kienlin et al., *The 2nd Fermi GBM Gamma-Ray Burst Catalogue : The First Four eras*, *ApJs*, **211** (2014) 13

[10]  M. V. Zombek, *Space Astronomy and Astrophysics*, *Handbook of Space Astronomy* (2009)

[11]  O. Godet et al., *Monte-Carlo simulations of the background of the coded-mask for X- and Gamma-raus on-board the Chinese-French GRB mission SVOM*, *NIMA*, **603** (2009) 265-371

[12]  A.Goldstein et al., *The Batse 5B Gama-ray Burst Spectral Catalog*, *ApJs*, **208** (2013) 21